\documentstyle[pra,aps,multicol,epsf]{revtex}

\begin{document}
\draft

\title{Worm algorithms for classical statistical models}

\author{ Nikolay Prokof'ev$^{1}$ and 
Boris Svistunov $^{1,2}$}

\address{
$^1$ Department of Physics, University of Massachusetts, 
Amherst, MA 01003, USA \\
$^2$ Russian Research Center ``Kurchatov Institute", 123182 Moscow, 
Russia}

\maketitle
\begin{abstract}
We show that high-temperature expansions may serve
as a basis for the novel approach to efficient Monte Carlo simulations. 
``Worm'' algorithms utilize the idea of
updating closed path configurations (produced by high-temperature expansions) 
through the motion of end points of a disconnected path.    
An amazing result is that local, Metropolis-type schemes 
may have dynamical critical exponents close to zero (i.e., their efficiency
is comparable to the best cluster methods). We demonstrate 
this by calculating finite size scaling of the autocorrelation time
for various universality classes.  
\end{abstract}

\begin{multicols}{2}
\narrowtext

Metropolis scheme  \cite{Metropolis}
is usually the most universal and  easy 
to program approach to Monte Carlo simulations. However,   
its advantages are virtually canceled out near phase transition points. 
It is believed that any scheme based on local \cite{2}  Metropolis-type
updates connecting system configurations into Markovian chain
is inefficient at the  transition point because 
its autocorrelation time, $\tau$, scales as $L^z$, where $L$ is
the system linear dimension and $z$ is the dynamical critical exponent
which is close to $2$ in most systems \cite{Yahata,Kawasaki}.  
 
An enormous acceleration of simulations at the critical point
has been achieved with the invention of cluster algorithms 
by Swendsen and Wang \cite{Swendsen}. However, the original method and its 
developments (both classical and quantum) \cite{Wolff,Evertz,Machta} are 
essentially non-local schemes, and we are not aware of any exception from
this rule.

In this Letter we propose a method which essentially 
eliminates the critical slowing down problem and yet remains local. 
The corner stone of our approach is 
the possibility to introduce  the configuration space
of closed paths. Closed-path (CP) configurations
may be then sampled very efficiently using Worm algorithm (WA)  
introduced in Ref.~\onlinecite{PRA} for quantum statistical models
in which closed trajectories naturally arise from imaginary-time evolution
of world lines. In classical models the CP representation derives
from high-temperature expansions for a broad class of lattice models (see, e.g.
Ref.~
\onlinecite{Parisi}). In 2D, another family of WA may be introduced
by considering domain-wall boundaries
as paths. 

We note, that our approach is based
on principles which  differ radically from cluster methods
and, most probably, has another range of applicability.
For one thing, the CP representation is 
most suitable for  the study of superfluid models by having  
direct Monte Carlo estimators for the superfluid stiffness 
(through the histogram of winding numbers \cite{Pollock})
which are not available in the standard site representation.  

In what follows we first recall how high-temperature expansions work
by employing Ising model as an example (still, trying to keep
notations as general as possible). We then explain
how WA is used to update the path configuration space. 
Next, we discuss specific implementations of WA for 
$|\psi |^4$-, {\it XY}-, and $q=3$ Potts models, and comment on
the special property of 2D models which allows an alternative
CP parameterization of the configuration space.    
The efficiency of the new method is studied
by looking at autocorrelation properties for six different 
universality classes. It is found that for 2D and 3D Ising models,
2D and 3D {\it XY}-models, and Gaussian
model the $\tau(L)$  scaling is consistent with the
law $\tau (L) = \tau_0+ c \ln (L) $,
i.e., its critical exponent is close to zero. For the two-dimensional 
$q=3$ Potts model our data are consistent with the power law with 
$z\approx 0.55$ which means that the Li-Sokal \cite{Li} bound $z>\alpha /\nu $
($\alpha$ and $ \nu$  are the specific heat and correlation
radius critical exponents)    
derived for the Swendsen-Wang algorithm is seemingly applicable to our method
as well. 

Since high-temperature expansions for various models 
can be found in standard texts (see, e.g., Ref.~\onlinecite{Parisi}) we
briefly remind the procedure for the Ising model
\begin{equation}
 -{H \over T} = \beta \sum_{b=<ij>} s_i\,s_j \;,
\label{H-Ising} 
\end{equation}
where $\beta =J/T$ is the dimensionless nearest-neighbor 
coupling parameter between spin variables $s_i=\pm 1$ and index 
$b=<ij>$ refers to the simple cubic/square lattice bonds
(we will also use another notation: $b=(i,\nu )$, in which
$\nu$ enumerates bonds containing site $i$). 
Since $H$ is additive, the corresponding Gibbs exponent 
factories in terms of exponents for each bond. Expanding 
each exponent in Taylor series  allows one to perform summation 
over site variables and to arrive at an expansion in powers 
of $\beta$. The partition function, for example, takes the form   
\begin{equation}
 Z=\sum_{\{s_i \}} \prod_{b=<ij>} 
 \left( \sum_{N_b=0}^{\infty} {\beta^{N_b} (s_is_j)^{N_b} \over N_b!} \right) 
   = \sum_{CP} W_{CP} \;.
\label{Z-Ising} 
\end{equation}   
Note that summation includes
all possible CPs, both connected and disconnected, with 
self-intersections and overlaps.
Graphically, each elementary factor in the order-by-order expansion
over the bond Hamiltonian can be represented by a line
ascribed to the corresponding bond \cite{Parisi}, 
and the closed path constraint is imposed by symmetry: the sum 
$\sum_{s_i} s_i^{k_i} $ is non-zero only if the number of bond lines 
connected to the site $i$ is even.  For future purposes 
it is convenient  to 
denote the number of bond lines $N_b=0,1,2, \dots$ and call it the ``bond state''. 
Then the ``site state'' is $k_i=\sum_{\nu} N_{i,\nu}$, 
and the configuration weight, $W_{CP}$, is given by           
\begin{equation}
W_{CP} = \left( \prod_b {\beta^{N_b} \over N_b!} \right) \, 
\left( \prod_i Q(k_i) \right)   \;,\;\;Q(k)= \sum_{s_i} s_i^k\;.
\label{W-Ising} 
\end{equation}  
[$Q(k)\equiv 2$ for the Ising model.]

\begin{figure}
\begin{center}
\epsfxsize=0.4\textwidth
\hspace*{0.5cm} \epsfbox{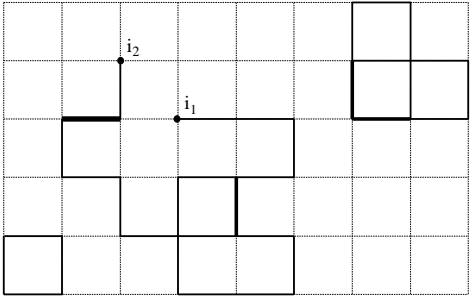}
\end{center}
\vspace*{-4.5cm}
\caption{A typical configuration with the ``Worm''. The two circles
correspond to the extra $s_{i_1}$ and $s_{i_2}$ variables, 
and the solid line width is proportional
to the bond state number $N_b$ (number of elementary lines);
0-th order terms are shown by dashed lines. 
}
\label{fig:fig1}
\end{figure}

The spin-spin correlation function $G(i_1-i_2)=\langle s_{i_1} s_{i_2}\rangle$
(we understand site indices as vectors)
by definition equals to $g(i_1-i_2)/Z$ where 
$g(i_1-i_2)=\sum_{\{s_i\}} s_{i_1} s_{i_2}\, e^{-H/T}$.
Configurations for $g$ differ from
those contributing to the partition function in only one respect: there are 
two special sites $i_1$ and $i_2$ with odd number of bond lines connected to them,
see Fig.~\ref{fig:fig1}. The corresponding site states are given by 
$k_{i}=1+\sum_{\nu} N_{i,\nu }$. These points are the only places where 
the path may be disconnected. We will abbreviate the configuration space
for $g(i)$ as CP$_g$, and note, that weights $W_{CP_g}$ are 
given by the same expression (\ref{W-Ising}).     

\vspace*{-0.6cm}

The above consideration is actually a perfect setup for the Monte Carlo
simulation since configurations for $Z$ and for $g(0)$ have identical 
bond elements (for the Ising model they even have equal weights, i.e. 
$g(0)=Z$).  Thus, if an ergodic  Metropolis process is  
sampling g-configurations according to their weights, then the Monte Carlo
estimators for $g(i)$ and $Z$ are just
\begin{equation}
g(i|CP_g)= \delta_{i_2-i_1,i}\;, ~~~~Z(CP_g)= \delta_{i_2-i_1,0} \;.
\label{Est-Ising} 
\end{equation} 
The spin susceptibility is given by 
$\chi = \sum_i g(i) / Z  $, and energy
may be computed in two ways: (i) as the nearest-neighbor sum 
$E=  (L^d /2) \beta \sum_{|i|=1} g(i) 
/ Z $, and (ii) using direct
estimator, $E={\cal E} /Z$,  which is non-zero only for the CP configurations
contributing to Z: 
\begin{equation} 
{\cal E}(CP)= \beta \sum_b N_b \;.      
\label{E-Ising} 
\end{equation}  
Our results 
for $\tau $ were obtained using this estimator.

The WA emerges as an idea to update $g$-configurations through the
space motion of the end points $i_1$ and $i_2$ only \cite{PRA}. The algorithm
itself is extremely simple and consists of two elementary updates
which are proposed in the context of a given configuration: 
if $i_1=i_2$, then
with probability $p_0$ we suggest to ``move'' both end points to site $i_0$
selected at random among all $L^d$ lattice cites, and with probability  
$p_1=1-p_0$ to ``shift''  the end point $i_1$ to one of the neighboring
sites by selecting the direction of the shift at random; if $i_1 \ne i_2$
we always suggest to ``shift'' $i_1$.

For the Ising model the move-update is automatically self-balanced
 and its acceptance ratio
is unity, otherwise it would involve the ratio of $Q$-functions
\begin{equation}
P_{\rm mv}(i \to j) = { Q(k_{j}+2) \over Q(k_{j})} \,{Q(k_{i}-2)
 \over  Q(k_{i})  }  \;.
\label{move} 
\end{equation}  
In contrast, shift-updates also increase/decrease the bond state by
``drawing/erasing'' bond lines, and  we select which way to proceed
with probability $1/2$. The shifts $i \to j$ with $N_{b=<ij>} \to N_b+1$ 
are balanced by shifts $j \to i$ with $N_b \to N_b-1$. 
By standard rules \cite{Metropolis}  we have acceptance ratios as  
\begin{eqnarray}
P_{\rm sh}(i \to j,N_b \to N_b +1) &=& r\,
{ \beta \over (N_b+1)} \; { Q(k_{j}+2)
 \over Q(k_{j}) }  \;, \\
\label{shift+} 
P_{\rm sh}(i \to j, N_b \to N_b-1) &=& r\,
{ N_b \over \beta }\; {Q(k_{i}-2) \over Q(k_{i}) }  \;,
\label{shift-} 
\end{eqnarray}   
where 
\begin{eqnarray}
         & 1/(2p_1)   &\mbox{~~if~} i_1=i_2=i \;, \nonumber  \\
 r\; = \;& 2p_1     &\mbox{~~if~} i_1=i,~i_2=j \;, \label{context} \\
         & 1        &\mbox{~~otherwise} \nonumber 
\end{eqnarray} 


The Ising model has a special property which can be used in 
practical applications to enhance the efficiency by truncating 
the configuration space. Namely,
using identity $e^{\beta s_is_j} \equiv \cosh \beta [1+\tanh \beta s_is_j ]$
one can restrict the bond summation to include terms $N_b=0,1$
only. The corresponding modifications of the scheme are straightforward.

We now turn to the $|\psi | ^4$-model described by the Hamiltonian
($\psi_i$ is a complex variable)
\begin{equation}
 -{H \over T} = \sum_{<ij>}\big( \psi_i \psi_j^* + c.c.\big)  
 + \sum_i \left[ \mu | \psi_i |^2 - U | \psi_i |^4 \right] \;, 
\label{H-psi} 
\end{equation}   
the two limiting cases of which are the Gaussian model, $U=0$, and
the {\it XY}-model, $U=\infty $.  
The procedure of factorizing the partition function
expansion in terms of the bond Hamiltonian is exactly as before.
The only new ingredient which is not
present in the Ising model, is that for each bond  we have two
{\it different} terms:  $\psi_i \psi_j^*$ and $\psi_i^* \psi_j$,
and therefore the expansion has to be performed for each of them separately.
Graphically, this can be captured by drawing lines with   
arrows, thus specifying each term by the arrow direction.
Correspondingly, the bond state is
defined by two numbers $(N_{b}^{(1)}, N_{b}^{(2)})$ 
which tell how many lines and in which direction 
go along this bond.
Since integrals 
$\int  d\psi  e^{ \mu  | \psi |^2  - U  | \psi |^4}
 \psi^m (\psi^*)^{m'} = \delta_{m,m'} Q(k=2m)$ are non-zero
only for $m=m'$, we conclude that the configuration space for $Z$
is defined on CP and site states are given by 
$k_i= \sum_{\nu} \left( N _{i, \nu}^{(1)} + N_{i,\nu }^{(2)}\right)=$even.
[The $Q(k)$ integrals are easily tabulated
prior to the simulation.] 

In close analogy with the previous case, the configuration
space for the two-point correlation function   
$g(i_1-i_2)= \int \prod_i\! d\psi_i \; \psi_{i_1} \psi_{i_2}^*\, e^{-H/T}$,
is given by CP$_g$ containing two special sites with
$k_{i}=1+\sum_{\nu}
\left( N _{i, \nu}^{(1)} + N_{i,\nu }^{(2)}\right)$.
WA updates consist of ``moves"  which change
the location of end points when $i_1=i_2$ and ``shifts'' of $i_1$ which
increase/decrease numbers $N _{i_1, \nu}^{(1)}$ and $N _{i_1, \nu}^{(2)}$.
Acceptance ratios are given by Eqs.~(\ref{move}-\ref{shift-}) 
where $N_b$ is understood as one of the numbers describing     
the bond state. An estimator for the winding number 
(which has exactly the same meaning as in worldline 
quantum Monte Carlo \cite{Pollock})
of the CP configuration is given by the difference between the 
bond numbers 
\begin{equation}
  M_{\alpha }= L^{-1} \sum_{b_{\alpha }} \left(
   N_{b_{\alpha }} ^{(1)} - N _{b_{\alpha }}^{(2)} \right) \;,
 \label{M} 
\end{equation} 
where $\alpha =x,y,z, \dots $, and notation $b_{\alpha} $ is used to
specify bonds connecting sites in direction $\alpha $.    
 
The $q=3$ Potts model (which we considered to compare with the exhaustive
cluster method study by Li and Sokal \cite{Li}) is described by
\begin{equation} -{H \over T} =
\beta \sum_{<ij>} (\delta_{s_i, s_j}-1)\;, \;\;\;
s_i=0,1,2\;.
 \label{H-Potts} 
\end{equation}   
Introducing phase variable $\varphi_i= (2\pi /3) s_i$   
we can rewrite Eq.~(\ref{H-Potts}) (up to a constant term) as
\begin{equation} -{H \over T} =
\beta ' \sum_{<ij>} \big( e^{i(\varphi_i - \varphi_j)}+c.c. \big)\;, \;\;\;
\beta ' = \beta /3 \;.
 \label{H-Potts1} 
\end{equation}    
With this form one may immediately proceed with the WA 
along the lines described above for the $| \psi |^4 $-model. However,
more efficient scheme results from the identity
$e^{2\beta ' \cos (\varphi_i - \varphi_j)} = A[1+\gamma  \cos (\varphi_i - \varphi_j) ]$, where $A=(e^{2\beta '}+2e^{-\beta '})/3$ and 
$\gamma = 2(e^{2\beta '}-e^{-\beta '})/3A$, which allows to restrict summation
over bond states to just three values $N_b=(0,0),~(1,0),~(0,1)$. 
 
We would like to comment that high-temperature expansions are 
{\it not} the only procedure to arrive at the CP configuration
space and WA. To underline this point we note that in lattice models
with discrete site variables one can unambiguously specify the state
of the system (up to symmetry transformations) 
by drawing domain boundaries, which, in 2D, may be considered as a CP
configuration. For the Ising model the configuration weight
is simply $W_{CP} = \prod_b e^{\beta (2N_b-1)} $ where the bond state 
takes values $N_b=0,1$.

To implement WA for efficient sampling of domain boundaries we 
formally enlarge 
the configuration space to include CP$_g$ configurations with two 
end points and proceed exactly as discussed above. The only difference is
that now ``open'' boundaries with $i_1 \ne i_2$ have {\it no} physical meaning
and serve just for updating purposes. Our results for the 2D Ising model
in Fig.~\ref{fig:fig2} were obtained using domain-wall representation.

\begin{figure}
\begin{center}
\epsfxsize=0.24\textwidth
\hspace*{0.0cm} \epsfbox{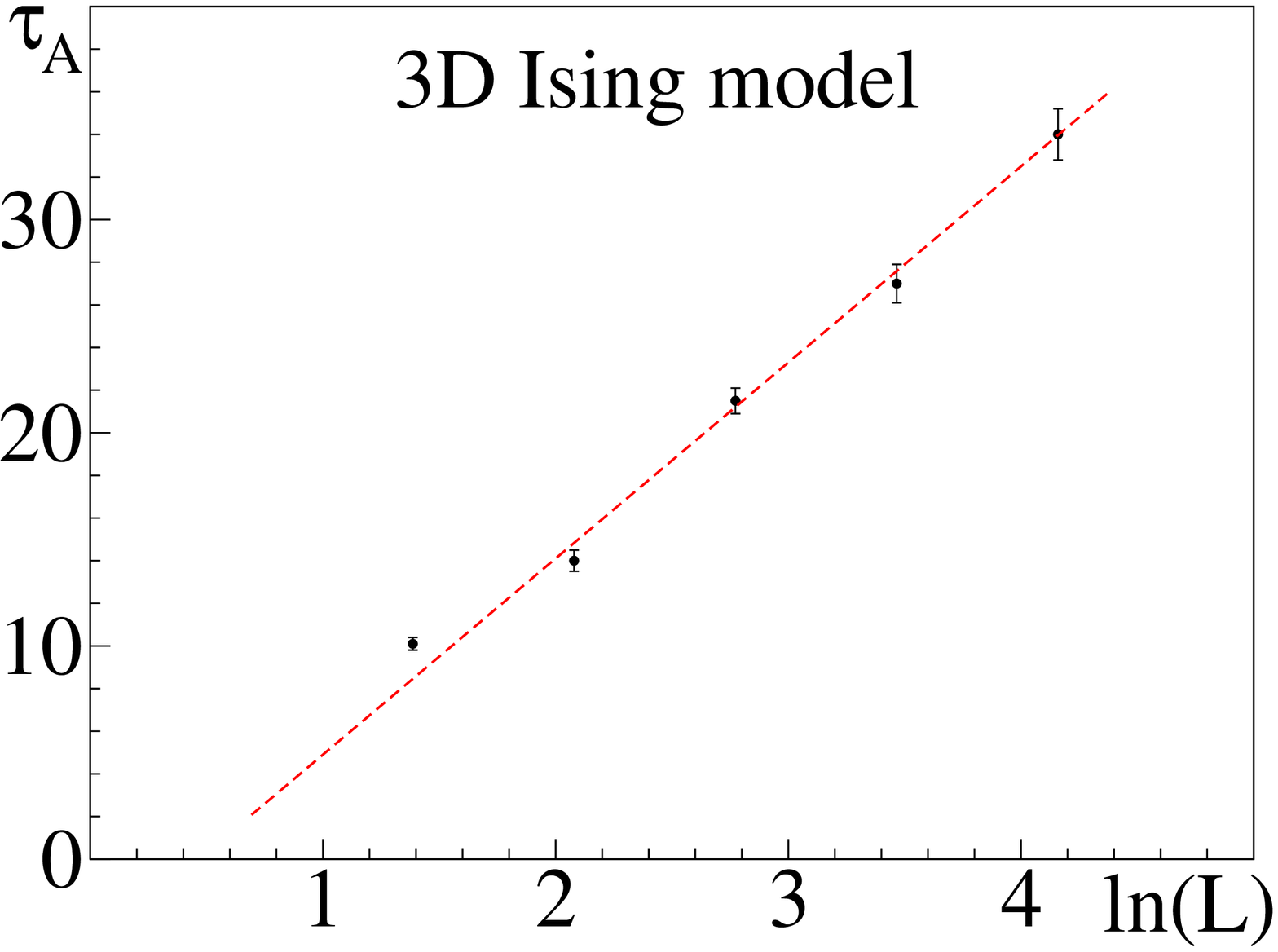}
\epsfxsize=0.24\textwidth
\hspace*{0.0cm} \epsfbox{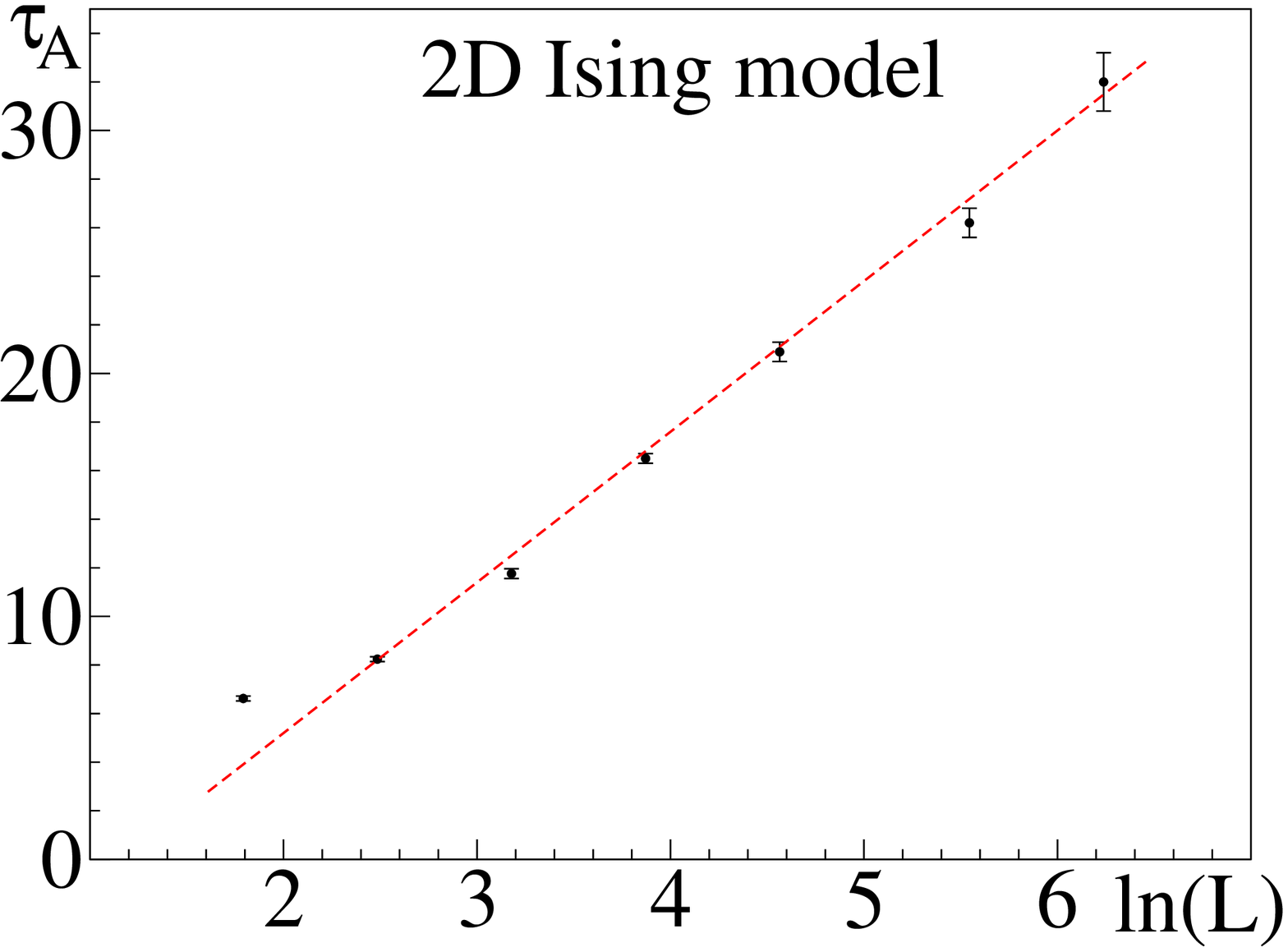}
\end{center}
\end{figure}
\vspace*{-3.5cm}
\begin{figure}
\begin{center}
\epsfxsize=0.24\textwidth
\hspace*{0.0cm} \epsfbox{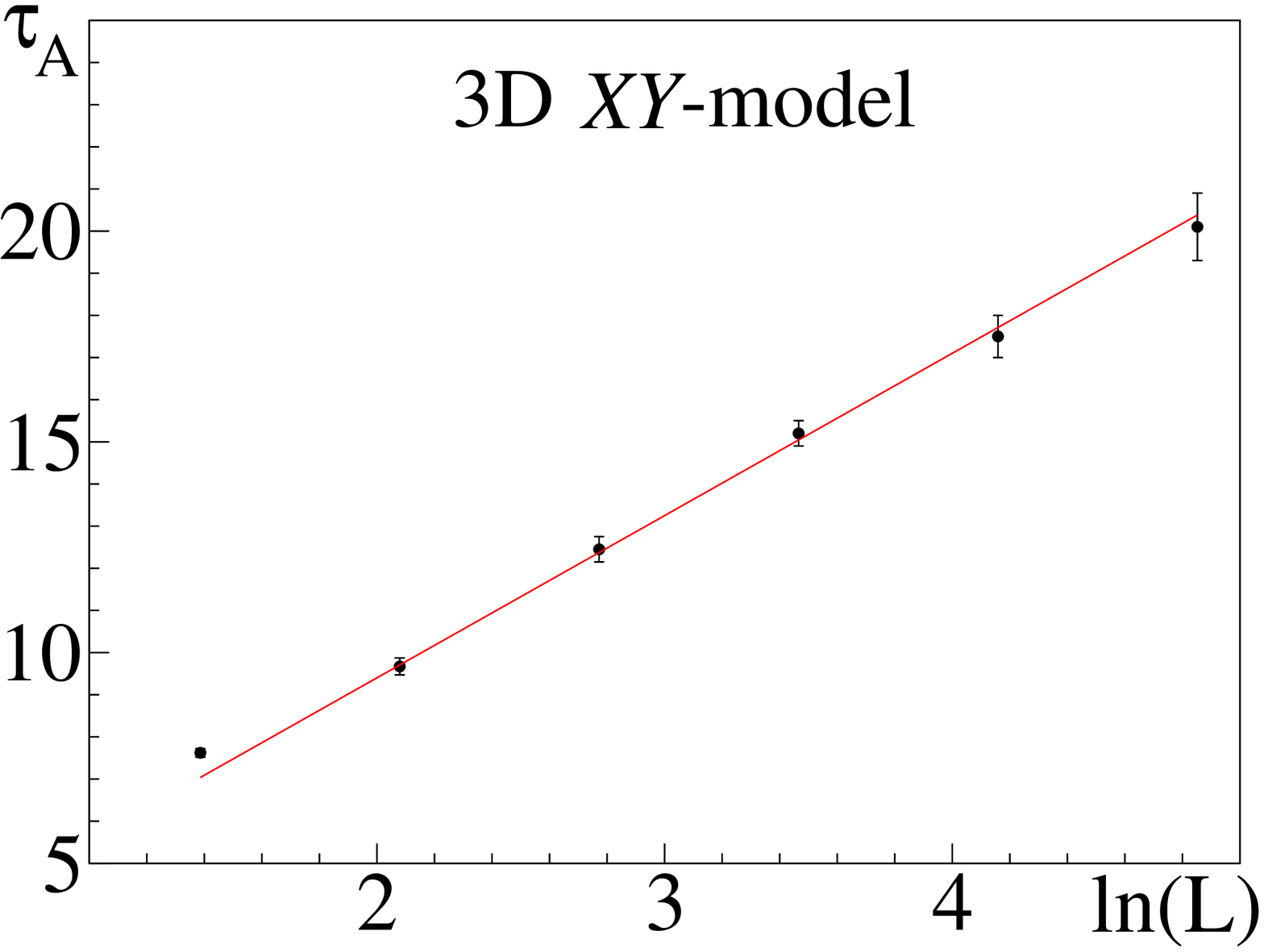}
\epsfxsize=0.24\textwidth
\hspace*{0.0cm} \epsfbox{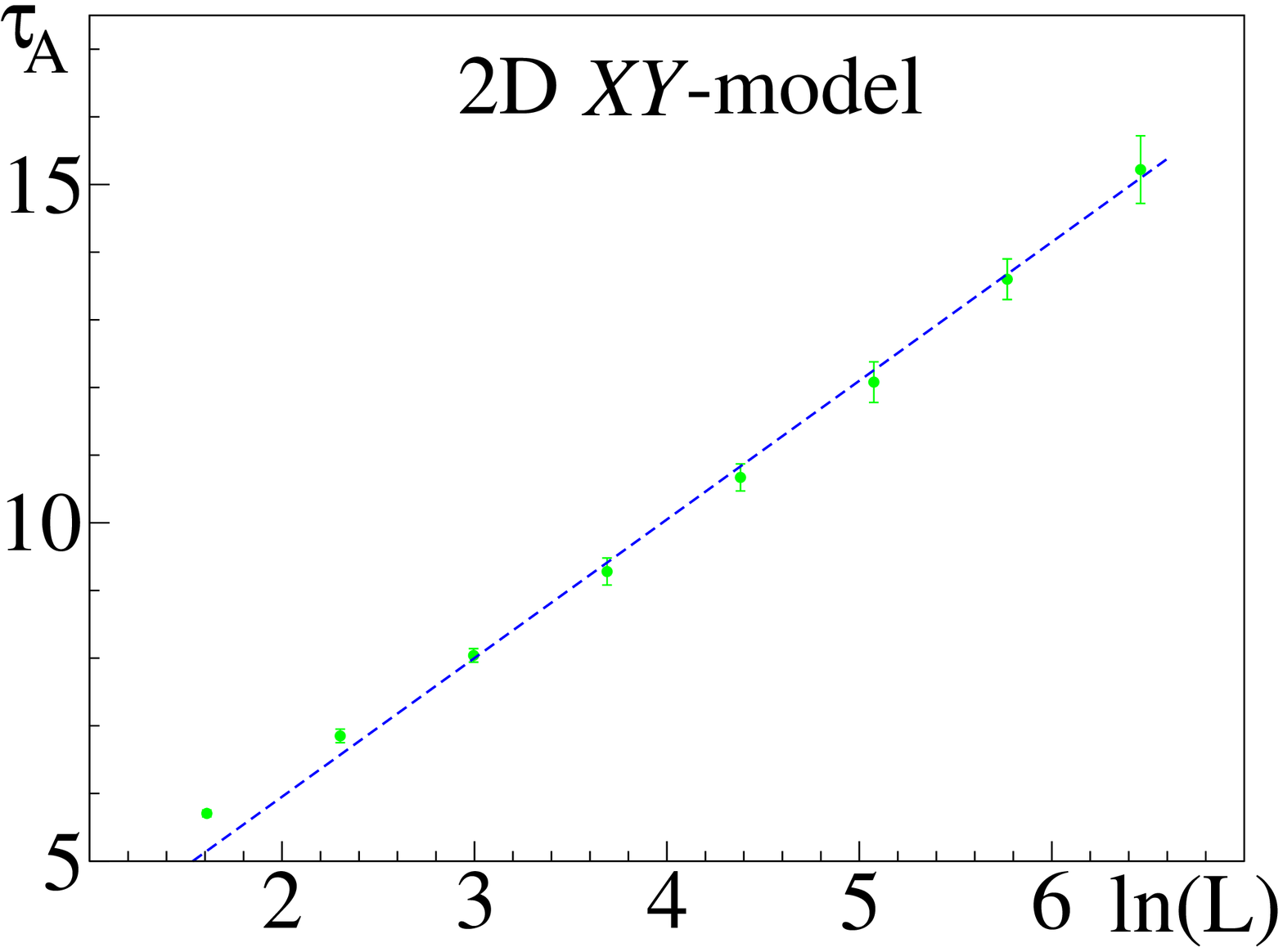}
\end{center}
\end{figure}
\vspace*{-3.5cm}
\begin{figure}
\begin{center}
\epsfxsize=0.24\textwidth
\hspace*{0.0cm} \epsfbox{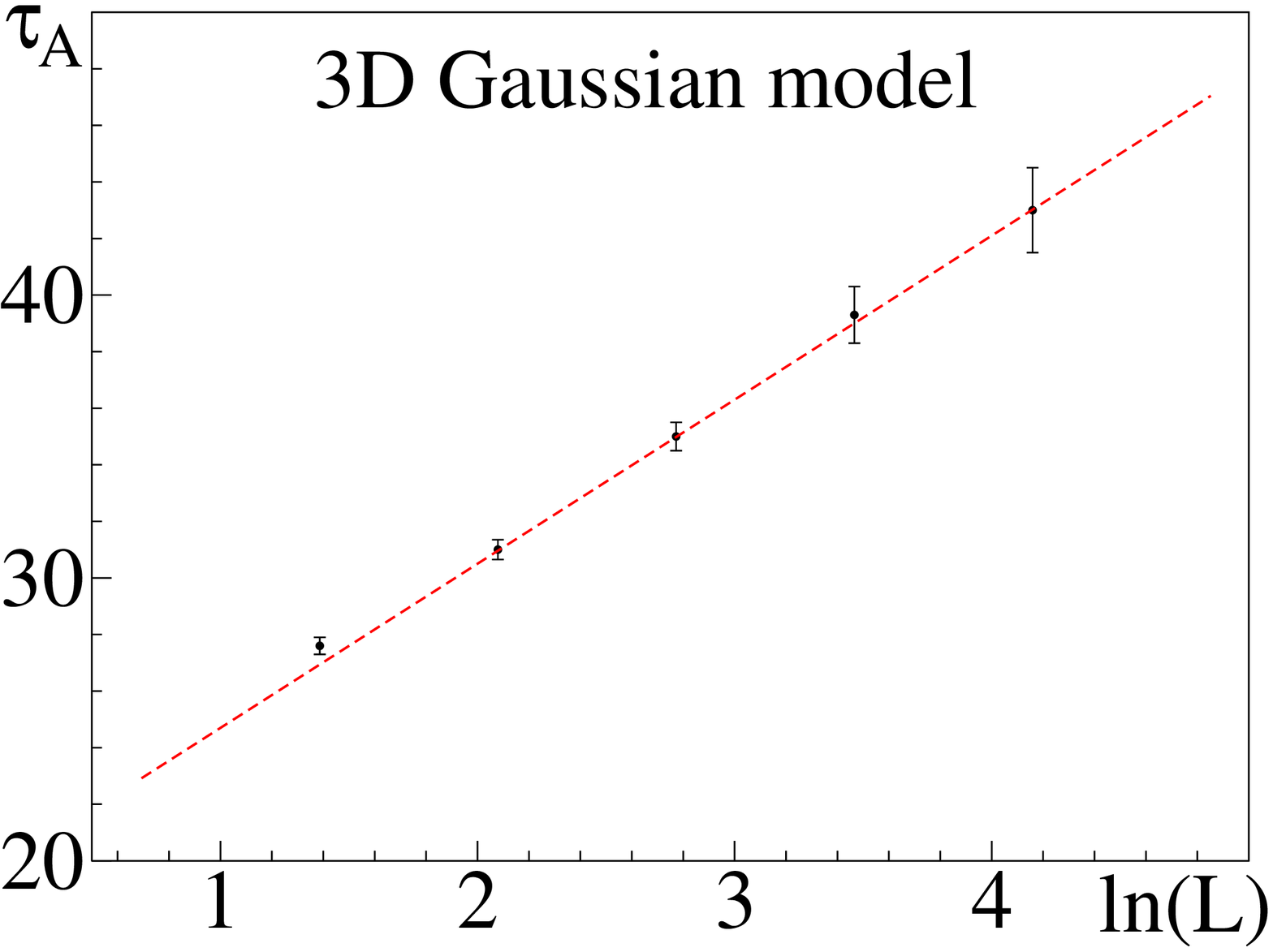}
\epsfxsize=0.24\textwidth
\hspace*{0.0cm} \epsfbox{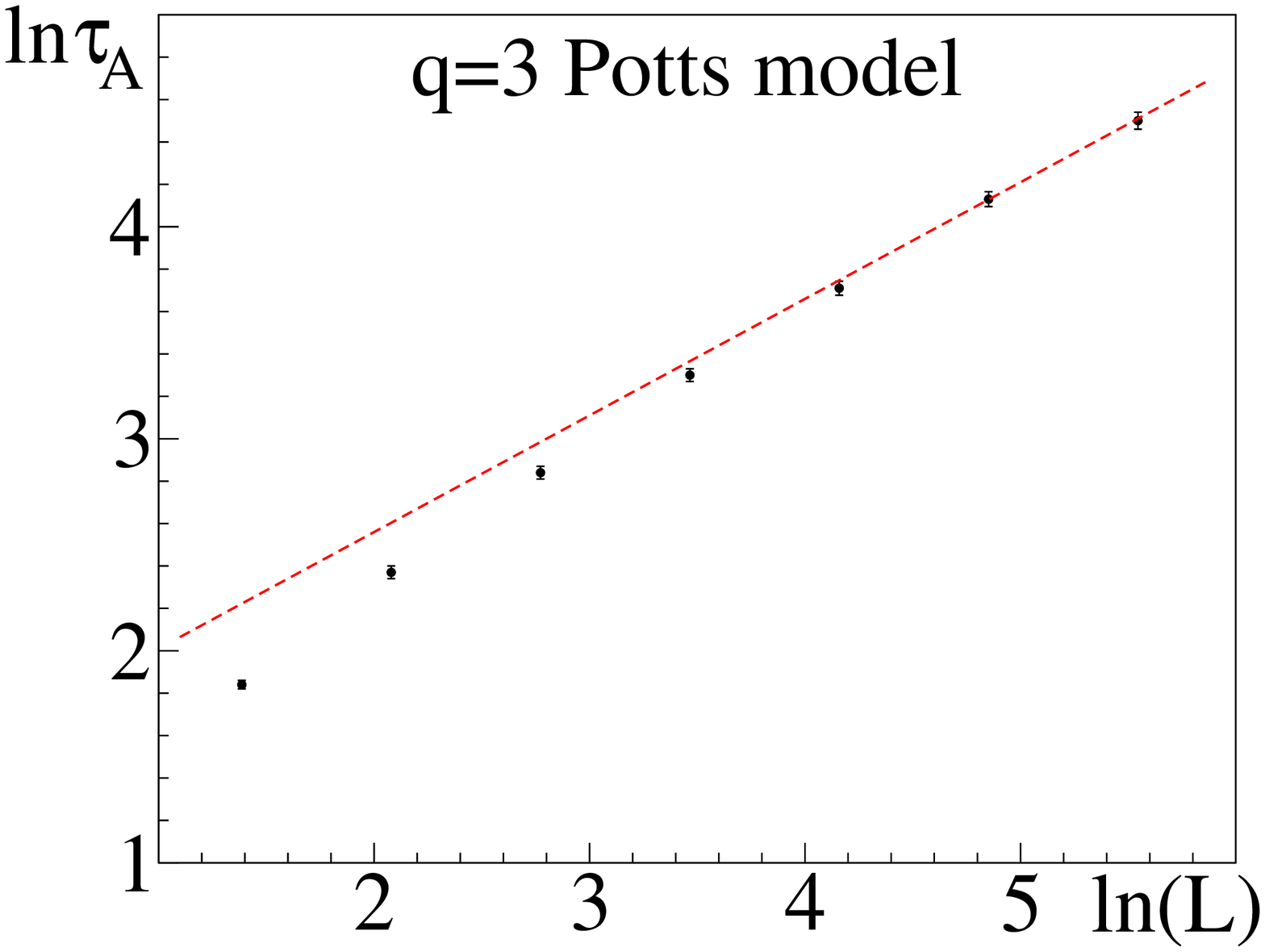}
\end{center}
\vspace*{-2.cm}
\caption{Autocorrelation times for various universality
classes. 
The 3D Ising model is
fitted to  $\tau = -4.3+9.2 \ln (L)$, and $z(L=64)=0.18$
(see text). 
The 2D Ising model is
fitted to  $\tau = -7.2+6.2 \ln (L)$, and $z(L=512)=0.25$.
The 3D {\it XY}-model is
fitted to $\tau = 1.7+3.85 \ln (L)$, and $z(L=128)=0.2$.  
The 2D {\it XY}-model
is fitted to $\tau = 1.85+2.05 \ln (L)$, and $z(L=640)=0.16$.
The 3D Gaussian model is
fitted to $\tau = 18.9+5.8 \ln (L)$, and $z(L=64)=0.17$.  
The $q=3$ Potts model in 2D
is fitted to the power law $\tau = 4.3 L^{0.55}$.
}
\label{fig:fig2}
\end{figure}

In Fig.~\ref{fig:fig2} we present data for the energy autocorrelation
time. In all cases, except $q=3$ Potts model in 2D, they are consistent 
with the linear in $\ln L$ behavior. Of course, we may not exclude that
logarithmic scaling is actually an intermediate length-scale behavior 
and at some larger $L$ it crosses over to the power law, $\tau (L) \sim L^z$, 
with small $z$. To get an estimate 
(upper bound) for the possible dynamic exponent 
we mention in the figure caption the slope $z(L)=d \ln (\tau )/d \ln (L)$
at the largest simulated $L$. For the $q=3$ Potts model our data
scale as $\tau \sim L^{0.55}$ for $L>64$, although the dynamic exponent
is showing a systematic decrease with $L$. This has to be compared with the 
value $z=0.515$ for the Swendsen-Wong algorithm \cite{Li}. It seems 
that the Li-Sokal lower bound for the dynamic exponent $z>\alpha /\nu $ also
applies to our method. For all models discussed here $ \alpha /\nu$ is very small
with exception of the $q=3$ Potts model were $\alpha /\nu =0.4$. 

To summarize, we found that WA Monte Carlo schemes working with closed path
representations of classical statistical models are as efficient as the best
cluster methods in reducing the problem of critical slowing down near
transition points.  The new approach is general enough to find applications
other than discussed in this Letter, and may be used to study quantities
like superfluid density for which other methods may not have direct estimators.

We thank J. Machta for valuable discussions.
This work was supported by the National Science Foundation under Grant
DMR-0071767.

\end{multicols}
        
\newpage       

\begin{thebibliography}{99}
\bibitem{2}
           Updates are local if they address configuration
           elements and change them without knowing about
           other elements outside the immediate neighbouhood
           of the updated part.
\end{thebibliography}
\end{document}